\documentclass[twocolumn,showpacs,preprintnumbers,amsmath,amssymb]{revtex4}

\usepackage{xspace}
\usepackage{graphicx}
\usepackage{dcolumn}
\usepackage{bm}

\def\beq{\begin{equation}}
\def\eeq{\end{equation}}

\def\bea{\begin{eqnarray}}
\def\eea{\end{eqnarray}}

\newcommand{\roughly}[1]%
    {{\mathrel{\raise.3ex\hbox{$#1$\kern-.75em\lower1ex\hbox{$\sim$}}}}}

\newcommand{\gsim}{\mathrel{\roughly>}}

\newcommand{\scr}[1]{\ensuremath{\mathcal{#1}}}

\newcommand{\Ga}{\ensuremath{\Gamma}}

\newcommand{\De}{\ensuremath{\Delta}}

\newcommand{\La}{\ensuremath{\Lambda}}

\newcommand{\si}{\ensuremath{\sigma}}

\newcommand{\om}{\ensuremath{\omega}}

\newcommand{\hc}{\ensuremath{\mbox{h.c.}}}

\newcommand{\TeV}{\ensuremath{\mathrm{~TeV}}}

\newcommand{\Eq}[1]{Eq.~(\ref{#1})}
\newcommand{\Eqs}[1]{Eqs.~(\ref{#1})}
\newcommand{\eq}[1]{(\ref{#1})}
\newcommand{\Ref}[1]{Ref.~\cite{#1}}
\newcommand{\Refs}[1]{Refs.~\cite{#1}}

\newcommand{\ttt}{\ensuremath{\bar{t}t}\xspace}
\newcommand{\ttb}{\ensuremath{\bar{t}b}\xspace}
\newcommand{\bbt}{\ensuremath{\bar{b}t}\xspace}

\newcommand{\Law}{\La_{\rm EW}}

\begin{document}


\title{Strong Electroweak Symmetry Breaking\\
and Spin 0 
Resonances}

\author{Jared Evans}
\email{jaevans@ucdavis.edu}
 \author{Markus A. Luty}%
 \email{luty@physics.ucdavis.edu}
\affiliation{%
Physics Department, University of California Davis\\
Davis, California 95616}%


\begin{abstract}
We argue that theories of strong electroweak symmetry breaking
sector necessarily contain new spin 0
states at the TeV scale in the
\ttt and \ttb/\bbt
channels, even if the third generation quarks are not composite
at the TeV scale.
These states couple sufficiently strongly to third generation
quarks to have significant production at LHC
via $gg \to \varphi^0$
or $gb \to t\varphi^-$.
The existence of narrow resonances in 
QCD suggests that the strong electroweak breaking sector
contains narrow resonances that decay to
\ttt or \ttb/\bbt, with significant branching fractions
to 3 or more longitudinal $W$ and $Z$ bosons.
These may give new ``smoking gun'' signals of strong electroweak
symmetry breaking.
\end{abstract}


\maketitle

\section{\label{sec:intro}Introduction}
One of the most important questions to be addressed at the
LHC is whether the physics that breaks electroweak symmetry
is strongly or weakly coupled.
Precision electroweak data are in good agreement with the
standard model with a light Higgs boson, but
mild cancellations may allow a good fit to precision
electroweak data in strongly-coupled models.
Direct searches are essential to settle this question.

One direct
test of the nature of electroweak symmetry breaking
is of course the search for the Higgs boson.
However, even if a light Higgs-like particle is discovered at the LHC,
it is important to make sure that it is ``the'' Higgs,
namely the state that unitarizes $VV$ scattering,
where $V = W,Z$.
There are other types of scalars
that naturally have couplings to gauge bosons
and fermions similar to Higgs bosons even though they
are not responsible for electroweak symmetry breaking,
for example radions \cite{radionHiggs} and
dilatons \cite{dilatonHiggs}.
In principle one can measure the
couplings of the scalar to electroweak gauge bosons
and compare with the values needed to unitarize $VV$
scattering, but this requires very high integrated luminosity
at LHC \cite{LHCHiggscoupling}.
Conversely, if the standard Higgs search does not lead to a discovery,
it does not follow that electroweak symmetry breaking is strongly
coupled.
For example, there may be a light Higgs with new physics modifying its
decays, making Higgs discovery difficult 
at LHC \cite{exoticHiggsdecay}.

It is therefore important to carry out direct searches
for a strongly-coupled electroweak symmetry breaking sector,
independently of the status of the search for the Higgs boson.
The classic signal is strong $VV$ scattering \cite{strongWW},
which is directly related to the absence of a light Higgs boson by
unitarity.
However, this also requires very high integrated
luminosity at LHC \cite{strongWWLHC}.

We argue that there is another generic signature
in models with strong $VV$ scattering:
new $J = 0$ states in the \ttt, \bbt, and \ttb channels,
 with masses of order a TeV.
These states must couple to the top quark sufficiently strongly
to change the \ttt, \bbt, and \ttb
cross sections by order 100\% at energies
of order TeV.

The existence of such resonances is already
expected in models where the top
quark is composite (as in ``topcolor'' models \cite{topcolor})
and extra dimensional models that are ``dual'' to strongly
coupled theories with a composite top quark \cite{RStop}.
We argue that such states also
exist in models where the top quark is an elementary particle
perturbatively coupled to a strong electroweak symmetry breaking
sector.
These states give rise to new signatures that may provide a
``smoking gun'' for strong electroweak symmetry breaking.

\section{\label{sec:unit}Strong Electroweak Breaking and the Top Quark}
We focus on models where the top quark mass arises from
coupling to an operator $\Phi$
with the quantum numbers of a Higgs doublet:
\beq\label{tHiggscoup}
\De\scr{L} = \frac{c}{\La_t^{d-1}}
\bar{Q}_L \Phi t_R + \hc
\eeq
Here $d$ is the scaling dimension of the operator $\Phi$
above the TeV scale,
and $\La_t$ is a mass scale that parameterizes the strength
of the coupling.
The dimensionless constant $c$ is chosen so that $\La_t$ is
the scale where this operator becomes strongly coupled (see below).
Another possibility not discussed here
is that the top quark couples to a fermionic
operator with quantum numbers conjugate to the
top itself \cite{KaplanTC}.
In order for the top to be weakly coupled to the electroweak
breaking sector at the TeV scale, we want $d$ to be as
small as possible, {\it e.g.\/}~$d = 1 + 1/\mbox{few}$.
On the other hand, naturalness requires that the operator
$\Phi^\dagger \Phi$ be irrelevant,
{\it i.e.\/}\ its dimension is larger than $4$.
The possibility of models satisfying these requirements
was pointed out in \Ref{CTC}.
Rigorous inequalities on dimensions in conformal field
theories allow this scenario \cite{RattazziCFT}.
Models based on QCD in the conformal window were described
in \Ref{CTClattice}.

The basic point is that the operator $\Phi$ creates
states in the strong sector, 
so \Eq{tHiggscoup} couples the top quark to the strong
sector.
As we now show, this coupling is sufficiently strong
to make an order 100\% change in the scattering
cross sections with initial states
$\bar{t}t$, $\bar{t}b$, and $\bar{b}t$
for $E \gsim \mbox{TeV}$,
where the electroweak symmetry breaking sector gets
strong.
For simplicity, we will discuss the electrically neutral
$\ttt$ channel below,
but the same arguments apply to the 
$\ttb$ and $\bbt$ channels.
For $m_t \ll E \ll \Law$ the chirality-violating
top quark scattering cross section
does not fall off at large energy
\cite{AppelquistChanowitz}:
\beq
\label{ttWWbelow}
\si(\bar{t}_L t_R \to VV) \sim \frac{1}{4\pi}
\frac{m_t^2}{v^4}.
\eeq
We now compare this to scattering amplitudes for $E \sim \Law$.
The leading contribution to the amplitude for 
chirality-violating top interactions involves an
insertion of the interaction \Eq{tHiggscoup}.
The cross section for producing a state $X$ with mass of
order $\Law$ is then of order
\beq
\label{ttabove}
\si(\bar{t}_L t_R \to X) \sim \frac{(4\pi)^3}{\Law^2}
\left( \frac{\Law}{\La_t} \right)^{2(d-1)}.
\eeq
The powers of $4\pi$ in this result can be understood 
from the fact that in the limit $\La_t \to \La_{\rm EW}$
the cross-section must be strongly coupled in the sense
of ``na\"\i{}ve dimensional analysis'' (NDA) \cite{NDA}.
To compare this with \Eq{ttWWbelow} we note that
NDA also gives
\beq
m_t \sim \Law \left( \frac{\Law}{\La_t} \right)^{d-1}
\eeq
and $\Law \sim 4\pi v$, so the cross sections in
\Eqs{ttWWbelow} and \eq{ttabove} are comparable.

We can also give an argument that does not rely on counting
factors of $4\pi$.
For $E \gg \Law$ the energy dependence of the total 
chirality-violating cross section to create
hadrons in the strongly coupled theory
is fixed by scale invariance:
\beq
\si(\bar{t}_L t_R \to \mbox{hadrons}) \sim 
\left( \frac{E^{d - 2}}{\La_t^{d - 1}} \right)^2.
\eeq
For $d \ne 2$, this has a different energy dependence
than at low energy (see \Eq{ttWWbelow}).
This means that the strong sector gives a correction 
to the cross section that is order 100\% at the 
matching scale $\Law \sim \mbox{TeV}$.
This argument does not imply a large change in the
cross section for $d \simeq 2$,
but the most phenomenologically interesting case is
$d < 2$, as discussed above.

The arguments above can be repeated for the chirality-violating
channels $\bar{t}_R b_L$ and $\bar{b}_L t_R$,
which also get contributions from the operator \Eq{tHiggscoup}.

\section{Resonances and Phenomenology}
We now discuss the nature of the new states 
in the $\ttt$, $\ttb$, and $\bbt$ channels at the TeV scale.
The most spectacular signals arise if these states include
narrow resonances.
NDA 
suggests that resonances
in a strongly-coupled theory with a scale $\Law$ will
have mass of order $\Law$ and width of order $\pi \Law$,
and will not be visible as individual resonances.
However, our experience with QCD suggests resonances
are present in strongly-coupled theories and are
significantly narrower than the NDA estimate.
Including the large-$N_c$ suppression of 
multi-meson couplings \cite{largeNQCD},
NDA gives
\beq
\left. \frac{\Ga(n\mbox{-body})}{m} \right|_{\rm NDA} 
\sim \frac{\pi}{N_c^{n - 1}}.
\eeq
This is for direct $n$-body decays,
{\it i.e.\/}\ those without 
intermediate on-shell particles.
In QCD we find much smaller widths \cite{PDG}:
\bea
\frac{\Ga(\rho \to \pi\pi)}{m_\rho} &\simeq& 0.2,
\\
\label{QCD3decay}
\frac{\Ga(\om \to \pi\pi\pi)}{m_\om} &\simeq& 10^{-2}.
\eea
The existence of narrow resonances is therefore
plausible even in a strongly coupled theory without
large $N_c$.

Because $\Phi$ is a Lorentz scalar, the
resonances created by the interaction \Eq{tHiggscoup}
are spin $0$.
The resonances will fall into representations of custodial
isospin symmetry, required to avoid large corrections to the
$T$ parameter.
Assuming the standard custodial symmetry breaking pattern
$SU(2)_L \times SU(2)_R \to SU(2)_C$,
the operator $\Phi$ transforms as
$(\bf{2}, \bf{2}) \to {\bf 3} \oplus {\bf 1}$.
We therefore expect isospin triplet and singlet states.

The mass of these states will be of order TeV.
The coupling to $\ttt$, $\ttb$, and $\bbt$ for these resonances
will be of order $y_t \sim 1$.
This coupling allows production of these states at LHC.
Electrically neutral states can be produced via
$gg \to \varphi$ via a top loop,
and electrically charged states can be produced via
$g b \to t \varphi^-$.
The production rate for these states at the LHC is shown in
Fig.~1.

\begin{figure}
\includegraphics{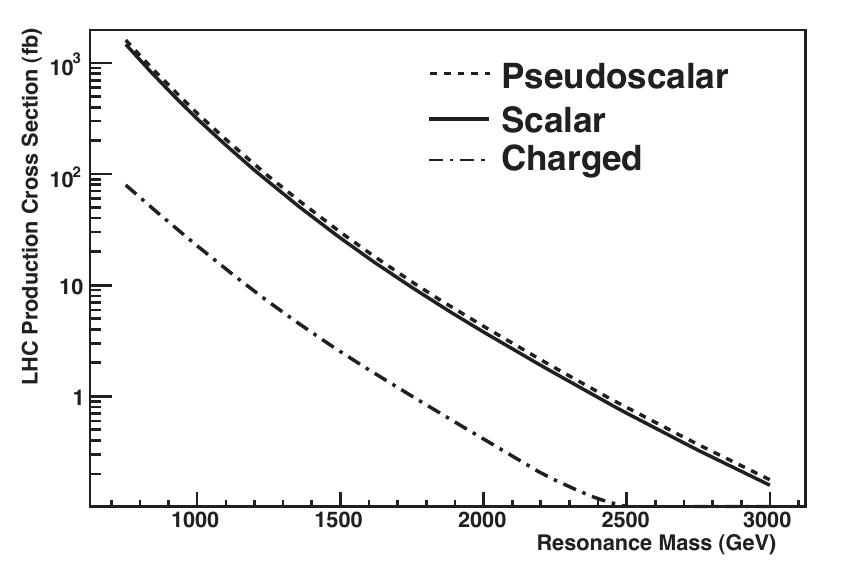}
\caption{\label{fig:epsart} Cross section for spin-0
resonance production at LHC via
$gg \to \varphi^0$ (solid and dashed lines)
and $gb \to t \varphi^-$
(dashed-dotted line).
We assume $g_{\ttt\varphi^0} = g_{\bbt\varphi^-} = y_t$.
Cross sections are tree-level with no $K$ factor.
Scalar and pseudoscalar cross
sections are nearly equal for the charged case,
and are not shown separately.}
\end{figure}

These resonances can decay to top quarks via
$\varphi^0 \to \ttt$ or $\varphi^- \to \bar{t} b$
via the coupling \Eq{tHiggscoup}.
For $m_\varphi \gg m_t$ we have
\beq
\frac{\Ga(\varphi \to \bar{t}t \mbox{ or } \bar{t} b)}{m_\varphi}
= \frac{3 (g_{\ttt \varphi,\,\ttb\varphi})^2}{16\pi}
\sim 10^{-1}.
\eeq
These resonances also have strong decays to longitudinal
$W$'s and $Z$'s, which are equivalent to the Nambu-Goldstone
bosons $\pi$ of the strong sector.
In the absence of additional symmetries,
isospin singlets will decay strongly to $\pi\pi$ with a large
width, similar to the TeV standard model Higgs.
A spin-$0$ isospin triplet cannot decay to $\pi\pi$,
so the leading strong decay is generally $\pi\pi\pi$.
If the 3-body strong decay is direct,
scaling from QCD gives $\Ga/m_\varphi \sim 10^{-2}$,
corresponding to a branching ratio of order 10\%.
Observation of a direct 3-body decay 
with such a large branching ratio is a
``smoking gun'' for strong dynamics,
since a perturbative 3-body decay would
have $\Ga/m \sim 10^{-4}$ due to 3-body phase space suppression.
Examination of the invariant mass distributions is required to
exclude the possibility of a 2-body chain decay.

Other interesting possibilities can arise if the strong
sector has additional discrete symmetries.
As an example, we consider a strong $SU(N)$ gauge theory in which the
operator $\Phi$ in \Eq{tHiggscoup} is a ``techniquark'' bilinear
$\Phi = \bar\psi_L \psi_R$.
The strong sector then preserves $C$ and $P$,
in which the lowest-lying resonances are
expected to have the quantum numbers of techniquark bilinears.
We can therefore classify the states by isospin, parity,
and $G$-parity ($G = C e^{i\pi I_2}$) as in QCD.
The operator $\Phi$ has the decomposition
\beq
I^{PG} = 0^{++} \oplus 0^{-+}
\oplus 1^{+-} \oplus 1^{--}
\eeq
so we may expect resonances in any of these channels.
We emphasize that a theory of this type does not necessarily
have QCD-like dynamics.
For example the theory may have additional
techniquarks that make the theory conformal above
the TeV scale \cite{CTClattice}.
Scalar resonances in QCD-like 
technicolor theories have been previously
discussed in \Refs{TCscalar}, but not the crucial role of the top
quark coupling in production.

The $0^{++}$ resonance is the analog of the QCD $\si$.
It has a 2-body strong decay to $\pi\pi$,
and is therefore expected to be broad.
Here $\pi$ is the composite eaten Nambu-Goldstone boson
that makes up the longitudinal polarization of the $W$ or $Z$.

The $0^{-+}$ is the analog of the QCD $\eta'$, and we call it 
the $\eta$.
Its most plausible strong decays are
$\eta \to \rho \pi\pi$ (followed by $\rho \to \pi\pi$)
or $\pi\pi\pi\pi$.
Here $\rho$ is the spin-1 $I^{PG} = 1^{++}$ particle
that is the analog of the QCD $\rho$.
The strong decay to $VVVV$ can plausibly compete with the
perturbative $\ttt$ decay, especially if 
the $\eta \to \rho\pi\pi$ decay is open,
leading to interesting observable signals at the LHC.
For example, assuming $\Ga / m_\eta \sim 10^{-2}$
for the strong decay, we obtain a cross section for
like-sign electrons or muons of order 1~fb for a
TeV resonance.
In QCD, $\eta \to \rho\pi\pi$ is kinematically forbidden,
but if even if we scale up QCD the decay is allowed
because
\beq
\frac{m_W}{m_\rho} 
\sim 10^{-1} \times
\left. \frac{m_\pi}{m_\rho} \right|_{\rm QCD}.
\eeq
This scaling also gives
$m_{\eta} \simeq 2.5\TeV$, $m_\rho \simeq 2\TeV$,
giving a very small production cross section.
However, this may be very misleading
because the dynamics is not expected to be QCD-like.

Another interesting case is the
$\pi'$ with $I^{PG} = 1^{--}$.
Its plausible strong decays are
$\pi' \to \pi\pi\pi$ or
$\pi' \to \rho\pi$ (followed by $\rho \to \pi\pi\pi$).
These possibilities correspond respectively either
to a narrow resonance with a possibly significant branching ratio
to $\pi\pi\pi$, or a broad resonance decaying dominantly
to $\pi\pi\pi$.
%
Finally, the $a_0$ with quantum numbers
$I^{PG} = 1^{+-}$ has plausible strong decays
$\eta\pi$ (followed by $\eta \to \ttt$), $\rho\pi\pi\pi$
(followed by $\rho \to \pi\pi$), and $\pi\pi\pi\pi\pi$.
The last two cases potentially give an observable
rate for a $VVVVV$ final state!

\section{Conclusions}
We have shown that the top quark coupling to strong electroweak
symmetry breaking provides a production mechanism for TeV-scale
spin-0 resonances.
We have argued that this is a generic signature
for strong electroweak symmetry breaking.
The processes $\ttt \to \varphi^0$ and $gb \to \varphi^- t$
may give significant numbers of events at the LHC, and can result
in the production of both narrow and broad resonances.
(By comparison, $WW$ scattering can only produce resonances
with 2-body strong decays, which are therefore broad.)
These resonances always have decays to third generation quarks
via $\varphi^0 \to \ttt$, $\varphi^- \to \ttb$,
but may also have substantial branching fractions
to multi-$W/Z$ final states.
Whether or not these modes are observable at the LHC
depends sensitively on their mass:
as can be seen from Fig.~1, the production cross section
drops by 3 orders of magnitude as the resonance mass
varies from $1$ to $3\TeV$.
However, besides strong $WW$ scattering this is the only
generic signal for strong electroweak symmetry breaking
observable at LHC, and should be pursued vigorously.
We do not know the masses of the lightest resonances, so broad-based
search strategies are required.
We leave the detailed investigation of phenomenology for future work.

\begin{acknowledgments}
We thank S. Chang, Z. Han, and J. Terning for comments on the manuscript.
M.A.L. thanks the Kavli Institute for Theoretical Physics
and the Aspen Center for Physics, where part of this work was
done.
\end{acknowledgments}


\begin{thebibliography}{99}

\bibitem{radionHiggs}
  C.~Csaki, M.~Graesser, L.~Randall and J.~Terning,
  Phys.\ Rev.\  D {\bf 62}, 045015 (2000)
  [arXiv:hep-ph/9911406];
  G.~F.~Giudice, R.~Rattazzi and J.~D.~Wells,
  Nucl.\ Phys.\  B {\bf 595}, 250 (2001)
  [arXiv:hep-ph/0002178].
  
\bibitem{dilatonHiggs}
  W.~D.~Goldberger, B.~Grinstein and W.~Skiba,
  Phys.\ Rev.\ Lett.\  {\bf 100}, 111802 (2008)
  [arXiv:0708.1463 [hep-ph]].
  
\bibitem{LHCHiggscoupling}
  D.~Zeppenfeld, R.~Kinnunen, A.~Nikitenko and E.~Richter-Was,
  Phys.\ Rev.\  D {\bf 62}, 013009 (2000)
  [arXiv:hep-ph/0002036];
  F.~Maltoni, D.~L.~Rainwater and S.~Willenbrock,
  Phys.\ Rev.\  D {\bf 66}, 034022 (2002)
  [arXiv:hep-ph/0202205];
  U.~Baur, T.~Plehn and D.~L.~Rainwater,
  Phys.\ Rev.\ Lett.\  {\bf 89}, 151801 (2002)
  [arXiv:hep-ph/0206024].

\bibitem{exoticHiggsdecay}
For a review, see
  S.~Chang, R.~Dermisek, J.~F.~Gunion and N.~Weiner,
  Ann.\ Rev.\ Nucl.\ Part.\ Sci.\  {\bf 58}, 75 (2008)
  [arXiv:0801.4554 [hep-ph]].
  
\bibitem{strongWW}
  B.~W.~Lee, C.~Quigg and H.~B.~Thacker,
  Phys.\ Rev.\  D {\bf 16}, 1519 (1977);
  M.~S.~Chanowitz and M.~K.~Gaillard,
  Nucl.\ Phys.\  B {\bf 261}, 379 (1985).
  
\bibitem{strongWWLHC}
  J.~Bagger {\it et al.},
  Phys.\ Rev.\  D {\bf 52}, 3878 (1995)
  [arXiv:hep-ph/9504426];
  J.~M.~Butterworth, B.~E.~Cox and J.~R.~Forshaw,
  Phys.\ Rev.\  D {\bf 65}, 096014 (2002)
  [arXiv:hep-ph/0201098].
  
\bibitem{topcolor}
  C.~T.~Hill,
  Phys.\ Lett.\  B {\bf 266}, 419 (1991).

\bibitem{RStop}
  K.~Agashe, A.~Delgado, M.~J.~May and R.~Sundrum,
  JHEP {\bf 0308}, 050 (2003)
  [arXiv:hep-ph/0308036];
  K.~Agashe, A.~Belyaev, T.~Krupovnickas, G.~Perez and J.~Virzi,
  Phys.\ Rev.\  D {\bf 77}, 015003 (2008)
  [arXiv:hep-ph/0612015].

\bibitem{KaplanTC}
  D.~B.~Kaplan,
  Nucl.\ Phys.\  B {\bf 365}, 259 (1991).

\bibitem{CTC}
  M.~A.~Luty and T.~Okui,
  JHEP {\bf 0609}, 070 (2006)
  [arXiv:hep-ph/0409274].

\bibitem{RattazziCFT}
  R.~Rattazzi, V.~S.~Rychkov, E.~Tonni and A.~Vichi,
  JHEP {\bf 0812}, 031 (2008)
  [arXiv:0807.0004 [hep-th]].
  
\bibitem{CTClattice}
  M.~A.~Luty,
  arXiv:0806.1235 [hep-ph].

\bibitem{AppelquistChanowitz}
  T.~Appelquist and M.~S.~Chanowitz,
  Phys.\ Rev.\ Lett.\  {\bf 59}, 2405 (1987)
  [Erratum-ibid.\  {\bf 60}, 1589 (1988)].


\bibitem{NDA}
  A.~Manohar and H.~Georgi,
  Nucl.\ Phys.\  B {\bf 234}, 189 (1984).

\bibitem{largeNQCD}
  G.~'t Hooft,
  Nucl.\ Phys.\  B {\bf 72}, 461 (1974).
  E.~Witten,
  Nucl.\ Phys.\  B {\bf 160} (1979) 57.

\bibitem{chargedHiggssearch}
For a review, see
  D.~P.~Roy,
  AIP Conf.\ Proc.\  {\bf 805}, 110 (2006)
  [arXiv:hep-ph/0510070].
  
\bibitem{PDG}
  C.~Amsler {\it et al.}  [Particle Data Group],
  Phys.\ Lett.\  B {\bf 667}, 1 (2008).
  
\bibitem{TCscalar}
P.~Di Vecchia and G.~Veneziano,
  Phys.\ Lett.\  B {\bf 95}, 247 (1980);
J.~Tandean,
  Phys.\ Rev.\  D {\bf 52}, 1398 (1995)
  [arXiv:hep-ph/9505256];
 
\end{thebibliography}
\end{document}